*Forum Südostasien / Forum South-East Asia*

# Customer Empowerment in Healthcare Organisations Through CRM 2.0: Survey Results from Brunei Tracking a Future Path in E-Health Research

Muhammad Anshari[1], Mohammad N. Almunawar[2], Patrick K. C. Low[3] & Zaw Wint[4]

## Introduction

The idea of customers' empowerment in healthcare emerged in response to the rising concern that customers or patients should be able to play a critical role in improving their own health. In traditional healthcare practices, a patient is the recipient of care as well as medical decisions. However, a paradigm shift has taken place – that is, a change from patients who merely receive care to those who actively participate in their healthcare. This change emerged in the 1960s and has spread throughout the entire healthcare industry as a social movement characterised by the right to act based on informed choice, active participation, a self-help perspective, and full engagement in critical processes (Kieffer, 1984).

Traditionally, healthcare providers have made most of the decisions on treatments. Indeed, the lack of participation of patients in healthcare processes was the main obstacle to the empowerment of patients as customers. Nevertheless, there will always be circumstances in which patients choose to hand over responsibility for

1  Muhammad Anshari is a PhD candidate at the Faculty of Business, Economics, and Policy Studies, Universiti Brunei Darussalam. Contact: anshari@yahoo.com (Corresponding author)

2  Mohammad N. Almunawar is a Senior Lecturer and a Deputy Dean of Faculty of Business, Economics, and Policy Studies, Universiti Brunei Darussalam. Contact: nabil.almunawar@ubd.edu.bn

3  Patrick K. C. Low is is a Professor and an Associate at the University of South Australia. Contact: patrick_low2003@yahoo.com

4  Zaw Wint is a Senior Lecturer at the PAPRSB Institute of Health and the Acting Director of Graduate Studies and Research Office, Universiti Brunei Darussalam, Brunei Darusssalam. Contact: zaw.wint@ubd.edu.bn

decisions about their healthcare to providers due to the difficulty in selecting available options or the time needed to understand the health problem and the options. However, this does not undermine the proposition that customers' empowerment will promote efficiency and that decisions should also be made from the perspective of customers (Segal, 1998).

The utilisation of information and communications technology (ICT), especially the Internet, in the healthcare sector is frequently referred to as electronic health or e-health. The main purpose of e-health is to improve healthcare management for the mutual benefit of patients and healthcare providers. One important aspect of e-health is how to manage relationships between a healthcare provider and its customers (patients) in order to create greater mutual understanding, trust, and patient involvement in decision making. Therefore, healthcare organisations are implementing Customer Relationship Management (CRM) as a strategy for managing interactions with their patients (Anshari & Almunawar, 2012). The recent Web technology (Web 2.0) facilitates customers to generate contents to accommodate both patient-health provider and patient-patient interactions. CRM enriched with the Web 2.0 (also called CRM 2.0) provides the capability for the intensive interactions mentioned above (Anshari, Almunawar, & Low, 2012). Besides, it can be considered a technology and strategy at the same time, raising hope for the advancement of e-health initiatives around the world.

The main goal of this paper is to introduce a promising future research direction which may shape the future of e-health systems. In this paper, we examine customers' expectations concerning the process of empowerment through CRM 2.0 to make customers more proficient in dealing with their own healthcare issues. A preliminary survey regarding empowerment in e-health services was conducted in Brunei Darussalam (Brunei). The Internet density in Brunei is high with over 75 percent of the population having access to the Internet, and there is a very narrow digital divide in Brunei (AITI, 2010). Internet literacy in Brunei is high, and so are people's expectations to be empowered through e-health. However, it has to be noted that the development of e-health in Brunei is still in its infancy. The future developments within the scope of e-health services in Brunei should therefore significantly focus on patients' empowerment as one of the important features required by the people. The rest of this paper is organised as follows: the following section presents the background of the study, followed by the research methodology. We then discuss the results of the study in relation to a CRM 2.0 model we propose based on this investigation.

## *Study Background*

Many researchers have discussed the issue of empowerment in healthcare organisations. For instance, empowerment can be analysed from the perspective of patient-healthcare provider interactions (Dijkstra, Braspenning, & Grol, 2002; Paterson, 2001; Skelton, 1997; van Dam, van der Horst, van den Borne, Ryckman, & Crebolder, 2003), or from the point of view of the patient alone (Anderson et al., 1995; Davison & Degner, 1997;; Desbiens et al., 1998; McCann & Weinman, 1996), or analysis can encompass both above mentioned perspectives (Golant, Altman, & Martin, 2003; Maliski, Clerkin, & Letwin, 2004; McWilliam et al., 1997). However, research that specifically discusses the issue of empowerment through CRM, and particularly CRM 2.0, is still very limited in the domain of e-health. Australia, a pioneer on this matter, has adopted a Personally Controlled Electronic Health Record (PCEHR) system, which stands out as an example for empowerment through e-health services (NEHTA, 2012). A significant element of patients' empowerment was achieved by allowing them to view their medical information electronically. However, PCEHR only enables patients to view their Electronic Health Record (EHR); it has not utilised features of CRM 2.0, which allow collaboration and conversation among patients or between patients and their healthcare providers. Therefore, our research primarily aims to develop a system that utilises features of CRM 2.0 in order to meet significant challenges in patients' empowerment in the domain of healthcare.

### *Customer Relationship Management*

The fact that customer expectations in healthcare services are high poses a serious challenge to healthcare organisations as they have to meet customers' requirements and make an exceptional impression on every customer (Anshari & Almunawar, 2011). CRM can be used by healthcare organisations both as a tool and a strategy to meet their customers' expectations. The terms of Social CRM and CRM 2.0 are used inter-changeably; both signify a new approach in CRM by allowing intensive interaction between customers and organisations and among customers by utilising Web 2.0. Hence, both share new distinct capabilities of social media and social networks that provide new approaches and potential which surpass the scope of traditional CRM. Cipriani (2008) described the fundamental changes offered by CRM 2.0 in terms of relationship, connection, and generated value.

The most significant feature of CRM 2.0 for e-health is the extensive network among customers and healthcare providers. This network is beneficial for customers as it provides multiple ways of interactive communication that enables sharing experience and knowledge electronically. Furthermore, it accommodates both the interaction between provider and patient, and between patient and patient. It also enables patients to generate contents of their personal health records.

With regard to these new features, Table 1 summarises the difference between CRM 2.0 and the precedent CRM 1.0 based on the type of relationship, connection, and generated value. CRM 1.0 mainly focuses on the individual one-way relationship, either customer to customer (C2C) or customer to business (C2B), whereas CRM 2.0 offers a collaborative relationship and thus supports the engagement in a more complex relationship network. Consequentially, CRM 2.0 enables customers to take a greater role in forming relationships. The connection types in CRM 1.0 present a limited view of the customer which has an adverse effect on their narrow information. On the other hand, CRM 2.0 enables multiple connections which allow customers to gain additional knowledge. In terms of generated value, CRM 1.0 is restricted to targeted messages, whereas CRM 2.0 offers more diverse value generation resulting from informal conversations of customers within social networks. A healthcare provider must consider the essential role of the conversation among patients as recent ICT enables them to share experiences without barriers. For example, the conversation that takes place among patients on a social network on healthcare services may influence the image of the healthcare provider. Shared experience of unprofessional treatments or services to patients can easily lead to distrust (Almunawar et al., 2012).

With all its features and benefits, CRM 2.0 can enhance the sense of control, improve self-care practices, and increase the scope of utilisations of healthcare services. The role of CRM 2.0 in empowering customers within healthcare organisations is further explored in the next section.

| Table 1: Comparison CRM 1.0 and CRM 2.0 | | |
|---|---|---|
| **TYPE** | **CRM 1.0** | **CRM 2.0** |
| **Relationship** | Focus on individual relationship (C2C, C2B) | Focus on collaborative relationship (engaging a more complex relationship network) |
| **Connection** | Limited view of the customer & his/her community preferences, habits, etc. | Multiple connections allow better understanding of the customer and his/her community |
| **Generated Value** | Targeted messages generate value | Conversation generates value |

Source: Cipriani, 2008

*Empowering Customers*

Empowerment is well supported in the healthcare literature and related to customers and healthcare services over the past decade (Dijkstra et al., 2002; Paterson, 2001; Skelton, 1997; van Dam et al., 2003). In the organisation, empowerment implies the provision of necessary tools to staff in order to be able to resolve, on the spot, most problems or questions faced by customers. Besides, staff can deal with customers directly and so reduce the number of dissatisfied customers who would otherwise have complained, but now simply switch brands (Low, 2002).

According to McWilliam et al. (1997), empowerment is a result of both interactive and personal processes, where caring relationships facilitate the emergence of *power* (or potential). Empowerment as an interactive process suggests that *power* is 'transferred' by one person to another, whereas empowerment as a personal process suggests that *power* is 'created' by and within the person. Although the expected outcome is the same, i.e. the gain of more *power* over one's life, the nature of the two processes is very different (Aujoulat et al, 2006). The first case entails that *power* can emerge through active co-creation and collaboration in an empowering relationship. Since CRM 2.0 facilitates interactions and collaborations, it can be used as a tool to implement empowerment. In the second case, when the process of empowerment is perceived from the point of view of the customers, it is considered as a process of personal transformation.

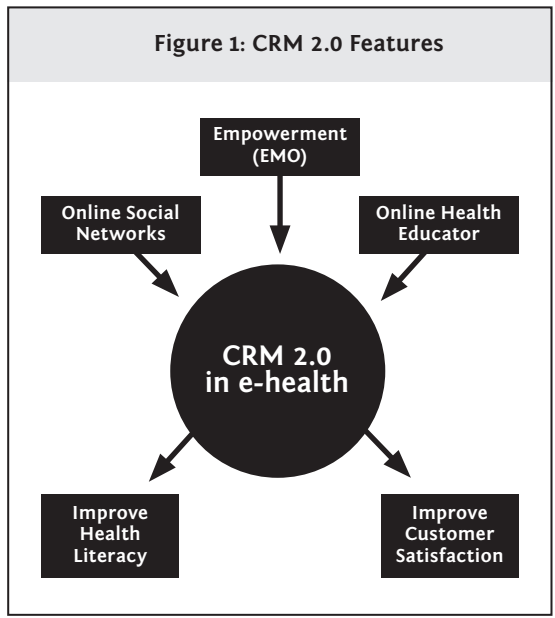

Figure 1: CRM 2.0 Features

Source: Authors' Compilation

Figure 1 presents three features of CRM 2.0 that can be embedded in e-health systems: online social networks, empowerment of electronic medical object (EMO), and online health educators. Online social networks present a facility for patient to patient interaction; EMO entails the ability of patients to access their specific medical record, which is empowered by healthcare organisations; online health educators are provided by a healthcare provider to facilitate the interaction between healthcare organisations and patients. Details of each component will be highlighted in the model discussion.

## Methodology

With all the promises and benefits offered by empowerment, the research attempts to utilise CRM 2.0 as both a tool and a strategy to empower customers in healthcare scenarios. For this purpose we single out potential features offered by CRM 2.0 to support customers' empowerment and prove these features in a survey. We use the purposive sampling methods in which respondents were intentionally selected from patients, patients' family, or medical staff from hospitals and homecare centres across the country. Questionnaires were developed based on the extracted features of CRM 2.0 (see Figure 1).

There were 366 respondents participating in the survey which was conducted from February to March 2011. Attributes of empowerment through CRM 2.0 were interrogated and analysed in order to gain preliminary responds from the potential users in Brunei. Data gathered from the survey was examined, interpreted and eventually converted in requirements to develop a prototype of CRM 2.0 for healthcare as shown in Figure 3. The prototype will be tested for the development of a system and for future recommendations.

| Table 2: Demographic Characteristics of Sample | |
|---|---|
| **Employment** | |
| 42% | Private |
| 58% | Government |
| **Gender** | |
| 46% | Male |
| 54% | Female |
| **Age** | |
| 13% | 20 years or younger |
| 38% | 21 - 30 |
| 19% | 31 - 40 |
| 18% | 41 - 50 |
| 12% | 51 years or older |
| **Education** | |
| 10% | Did not complete high school |
| 31% | Completed high school only |
| 59% | Completed more than high school |
| **Internet Usage** | |
| 63% | At least daily |
| 18% | Daily to weekly |
| 9% | Weekly to monthly |

## Analysis and Discussion

To analyse the reliability of the questionnaire items used in this study, Cronbach's alpha is used to measure internal consistency (1). Cronbach's Alpha[5] measured 0.80 for 363 items which indicates a relatively high internal consistency and therefore reliability of the study. Table 2 shows the demographic characteristics of the samples. Important to note here is that respondents between the age of 20 to 50 years are the most potential users of CRM 2.0 notably because of their basic Internet literacy which is the critical success factor for the empowerment through CRM 2.0.

$$\alpha = \frac{N.\bar{c}}{\bar{v} + (N - 1).\bar{c}} \quad (1)$$

Table 3 below shows the results of the survey in percentage. Interestingly, the survey reveals that the expectations of customers toward empowerment through CRM 2.0 are very promising.

| Table 3: Empowerment Features in CRM 2.0 | |
|---|---|
| **Empowerment of EMO** | |
| 83% | Request appointment online |
| 79% | View Electronic Medical Record (EMR) |
| 69% | Update personal medical activities |
| 81% | View health promotion online |
| 76% | Record health activities online |
| 84% | View payment online |
| **Online Health Educator** | |
| 77% | Prescription online |
| 78% | Referral online |
| 50% | Paying service online |
| 73% | Consultation online |
| 92% | EMR only for trusted doctors |
| **Social Networks** | |
| 62% | Discuss health service in social network |
| 77% | Control of own EMR |
| 72% | Discuss health status in social network |
| 76% | Discuss with patients same condition |

Source (Both Tables): Authors' Survey Results

---

5 Cronbach's alpha determines the internal consistency or average correlation of items in a survey instrument to gauge its reliability.

*The Proposed CRM 2.0 Model*

Based on the literature review and the survey results, we propose a CRM 2.0 model to accommodate the features of patients' empowerment. CRM functionalities are composed of Marketing, Sales, and Customer Service, which are operated to achieve the business strategy of a healthcare organisation. For example, a marketing strategy should accommodate social marketing to promote public health and commercial marketing to acquire more customers coming for services. Customer service will offer distinct value for each activity. CRM 2.0 accommodates various features and components of empowerment in healthcare systems, as its central role entails self-managed data and authorisation to encourage customers to provide full information in relation to their health. This is very important to healthcare organisations as it also enables customers to access more information especially on EMO. Organisations will benefit from the strategy of marketing, sales, and customer service, while customers will benefit from an empowerment which promises better customer satisfaction and health literacy.

Figure 2 presents a proposed model of CRM 2.0 in healthcare organisations. It offers a starting point for identifying possible theoretical mechanisms that might account for ways in which CRM 2.0 provides one-stop service for building relationships between a healthcare organisation, patients, and community at large. The framework is developed from Enterprise Social Networks, Internal Social Networks, listening tool interfaces, and healthcare value configuration.

The term social networks refers to any Web 2.0 technology accessed by patients or their families. We focus on two different types of interconnection: Enterprises Social Networks and Internal Social Networks. The Enterprises Social Networks refers to external and popular Web 2.0 applications such as *Facebook, Twitter, LinkedIn, MySpace, Friendster* etc. which may serve as platforms for interaction between patients. The dashed line connecting Enterprises Social Networks and CRM systems means that none of these networks has control over the other directly. However, constructive conversation and information from Enterprises Social Networks should be captured for creating strategies, innovations, better services and, at the same time, respond accurately to emerging challenges. Furthermore, the model involves Internal Social Networks that are operated, managed, and maintained within the healthcare organi-sation's infrastructure. The pivotal target is a conversation between internal patients and their families with regard to similar health problems and illness. For example, patients with diabetes will be motivated to share their experience, process of learning, and knowledge with other patients.

In general, the aim for having external and internal social networks is to engage patients and export ideas, foster innovations of new services, and ensure quick response/feedback for existing services and technologies from people inside and outside the organisation. Both provide a range of roles for the patient or his/her family; this is in line with the survey results presented in Table 3.

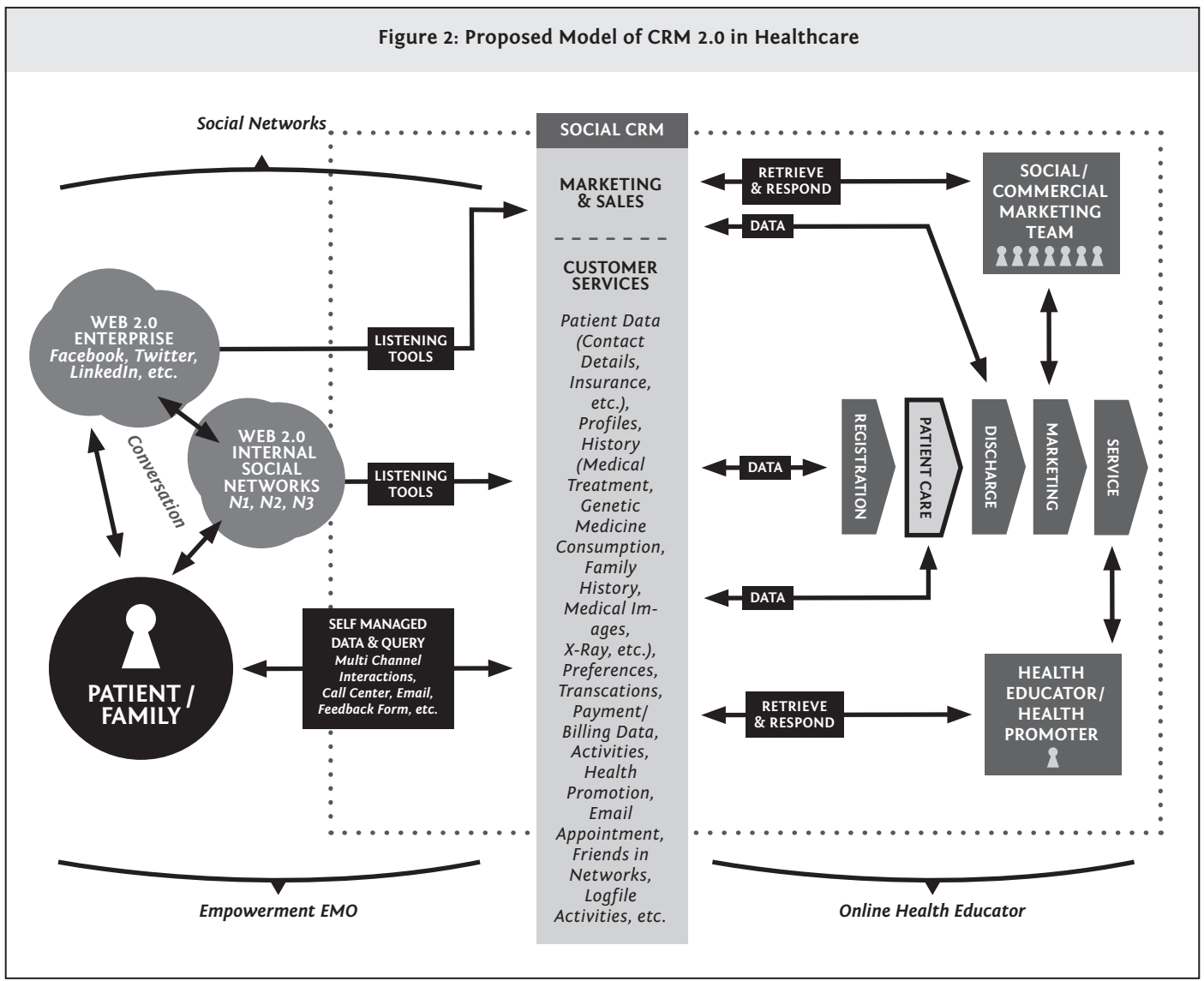

**Figure 2: Proposed Model of CRM 2.0 in Healthcare**

Source: Authors' Compilation

CRM 2.0 empowers patients (and their family) to control their own data. Once the patient registers for the service from the healthcare provider, it will enable him/her to enjoy the benefits of a personalised e-health system with CRM 2.0 as the frontline of the system. Authorisation will be provided for each patient. Hence, the authorisation and self-managed account/service grant the access to all applications and data offered by the system. Technical assistance is provided by the manual or by a health informatics officer (just like any other customer service in business/organisation), who is available online and assists patients and their family in utilising the system. Furthermore, since all information (medical record) can be accessed online anywhere and at any time, it can contribute to a collaborative treatment in telemedicine. This feature refers to the empowerment through EMO (see Figure 1 and Table 3).

The model adopts a modular approach; it will assist a healthcare provider to initiate empowerment by stages, and measure the performance gradually. Some of the features available to the users include the update of personal data, access to medical records and history (medical treatment received, medicine consumption history, family illness history, genetic, medical imaging, x-ray and so forth), preference services, transaction, payment/billing data, activities, personal health promotion and education, e-mail, appointment, friend in networks, forums, chatting and so forth.

Finally, a healthcare business scenario is a critical process that affects personal health as much as it affects healthcare organisations. It is important for healthcare organisations to ensure that CRM 2.0 is fully utilised by their customers. Patients need to collaborate with healthcare providers in order to gain sufficient know-how to use electronic and online services effectively. To support this function, we propose an Online Health Educator which enables patients to attain a better knowledge and control over their health data, and yet, contributes to the basic communication between patients and healthcare providers. Additionally, the Online Health Educator determines the success of the implementation as it ensures that there is a group of staff dedicated to guarantee that e-health services are managed in a professional way.

*Conclusion*

Patient expectations in healthcare are high, and this creates a serious challenge for healthcare providers. In this article, we suggest a model to address patients' demand by combining the concept of value chain and value configuration in healthcare organisations, CRM, and Social Networks. The model incorporates customer empowerment through healthcare organisation to patient as well as patient to patient relationships. Recent Web technology offers a broader outlook on customers' empowerment in many ways and levels depend on the needs and policies of the organisations. These range from digitalising medical records to the customers' ability in accessing their EHR.

We conducted a survey to verify our empowerment features established in the CRM 2.0 model. The result of the survey confirms that customers prefer the empowerment features deriving from the model. They prefer to have control over information on their health and over applications that may affect their health. Moreover, it also confirms that the availability of an online health educator proposed in the model is important in order to achieve the goals of e-health in educating and promoting better health to customers. CRM 2.0 shares the exceptional capabilities of social media and social networks that provide a new approach which transcends traditional CRM. The majority of respondents agreed that both social networks and social support online should be part of an e-health system.

The future direction of the study is to design a prototype based on our model. The prototype will be used to validate our model and to confirm that the empowerment featured in an e-health system is essential and highly recommended.